\documentclass[graybox,natbib,nosecnum]{svmult}
\bibpunct{(}{)}{;}{a}{}{,} 

\pdfoutput=1   

\usepackage{mathptmx}       
\usepackage{helvet}         
\usepackage{courier}        
\usepackage{type1cm}        

\usepackage{makeidx}         
\usepackage{graphicx}        
\usepackage{multicol}        
\usepackage[bottom]{footmisc}
\usepackage[normalem]{ulem}	
\usepackage{hyperref}  

\usepackage{soul}   

\newcommand*\aap{A\&A}

\newcommand*\aj{AJ}

\newcommand*\apj{ApJ}
\newcommand*\apjl{ApJ}

\newcommand*\apjs{ApJS}

\newcommand*\nat{Nature}

\newcommand*\pasp{PASP}

\newcommand*\procspie{Proc SPIE}

\newcommand{\hbindex}[1]{{#1}}  

\makeindex             

\begin{document}

\title*{Radial Velocities as an Exoplanet Discovery Method}
\author{Jason T.\ Wright}
\institute{Department of Astronomy \& Astrophysics and 
Center for Exoplanets and Habitable Worlds 525 Davey Laboratory 
The Pennsylvania State University 
University Park, PA, 16802, USA \email{astrowright@gmail.com}}
%
%
\maketitle

\abstract{The precise radial velocity technique is a
  cornerstone of exoplanetary astronomy. Astronomers measure Doppler shifts in the
  star's spectral features, which track the line-of-sight
  gravitational accelerations of a star 
  caused by the planets orbiting it. The method has its roots in
  binary star astronomy, and exoplanet detection represents the
  low-companion-mass limit of that application. This limit requires
  control of several effects of much greater magnitude than the signal
  sought: the motion of the telescope must be subtracted, the
  instrument must be calibrated, and spurious Doppler shifts
  ``jitter'' must be mitigated or corrected.  Two primary forms of
  instrumental calibration are the stable spectrograph and absorption
  cell methods, the former being the path taken for the next
  generation of spectrographs. Spurious, apparent Doppler shifts
  due to non-center-of-mass motion (jitter) can be the result of
  stellar magnetic activity or photospheric motions and granulation.
  Several avoidance, mitigation, and correction strategies exist,
  including careful analysis of line shapes and radial velocity
  wavelength dependence. }

\section{Radial Velocities in Astronomy}

Most of what we know about heavens comes from information encoded in
various forms of light. While many important advances have come from studies of
meteorites, {\it in situ} measurements of the Solar System by space
probes, detection of high energy particles, and more recently by
observations of astronomical neutrinos and gravitational waves; the
vast majority of astronomical observations come from eking as much
information as possible from photons.  

Because of this restriction, certain properties of stars and galaxies are much more
easily determined than others. Positions on the sky can be measured
to great accuracy, but distances can often only be approximated.
For the time derivatives of these quantities, the situation is usually reversed; motions of objects in the plane of the sky can be
imperceptible due to the vast distances involved, but thanks to the Doppler shift motions along the line
of sight (\hbindex{radial velocities}) can often be measured with great
precision.

Much of astronomy is concerned with the motions of objects under the
influence of their mutual gravity, and so radial velocity measurements
naturally form a cornerstone of astronomical research. They tell us
the masses of everything from moons in the solar system to stars to
galaxies; they have revealed the existence
of unseen planets, dark matter, and the expansion and
acceleration of the universe itself.

\subsection{Binary Systems}

Binary stars provide much of our foundational knowledge of stellar
structure. The relative masses of the stars in a binary system can be determined from the stars'
orbits about their common center of mass from the ratio of the
amplitudes of their radial velocity variations. Further knowledge of
the non-radial component of their motion --- either from astrometric
measurements of their orbits or from the fact that they eclipse each
other --- yields absolute masses for the individual stars from
Newton's Laws.  

In some cases, only a single star in a binary system is bright enough
to have its spectrum measured (a \hbindex{single-lined spectroscopic
  binary} system, or SB1). In this case, the (semi-)amplitude of the radial
velocity variations $K$ over the course of an orbit with period $P$ is related to the
properties of the system via the so-called \hbindex{mass function}, $f$:

\begin{equation}
f \equiv \frac {P K^3  (1-e^2)^{\frac{3}{2}}}{2\pi G}= \frac{M_{\rm
    unseen}^3\sin^3i}{(M_{\rm unseen}+M_{\rm seen})^2}
\label{massfunction}
\end{equation}

\noindent Here, $e$ represents the eccentricity of the orbit, $i$
represents the inclination of the plane of the orbit with respect to
the plane of the sky ($i=0$ represents a face-on orbit with no radial
component), $G$ is Newton's constant, and $M_{\rm seen}$ and $M_{\rm unseen}$ are the
masses of the seen and unseen stars.  The phase and orientation of the
elliptical orbit with respect to the line-of-sight are represented by
an additional pair of parameters, $T_0$ and $\omega$, and the overall
radial velocity of the center of mass of the system is usually given by
$\gamma$.  The orientation of the orbit on the plane of the sky is
given by $\Omega$ but, as with $i$, its measurement requires
additional information not found in the radial velocities.  (For a
more detailed discussion of single-lined spectroscopic orbits in
general, see 
\citet{Wright_Gaudi}, and for a more detailed treatment of the
mechanics of translating measurements into orbital parameters see
\citet{Wright09b}.)

The orbital parameters measured in this way represent the orbit of the
{\it observed} star about the center of mass of the system; in
single-lined system the existence of the unseen object is often {\it
  inferred} from this motion, and its (unmeasured and usually
unreported) orbital elements are identical except that $K$ is scaled
by a factor of $M_{\rm unseen}/M_{\rm seen}$ and $\omega$ differs by
$\pi$. 

This application of radial velocities is not restricted to
binary stars; it is used to detect unseen
binary companions ranging in mass from black holes to planets.  For
the case of exoplanets, we can approximate the mass function in the large mass-ratio
limit $M_{\rm unseen}/M_{\rm seen} = M_{\rm planet}/M_* \ll 1$, yielding the more familiar
equation relating the amplitude of a Doppler shift to the mass and
orbital properties of the planet:
\begin{equation}
K \approx \left( \frac{2\pi G}{P M_*^2} \right)^\frac{1}{3} \frac{M_{\rm
    planet} \sin{i}}{\sqrt{1-e^2} }
\end{equation}

The ability to detect exoplanets with precise radial velocimetry thus
depends on the planet's mass, inclination, and orbital period (there
is also a dependence on orbital eccentricity, which is complex but weak for
low eccentricities).  The period dependence is weak, and for giant
planets, detectability is often limited more by the duration of the observations than
their RV amplitude.  

The strongest dependence is on the quantity $M_{\rm planet} \sin{i}$ (pronounced ``em-sine-eye'').  The true mass of the planet is larger by a
  geometric factor $1/\sin{i}$, and so this is often 
  referred to as the ``minimum mass'' of the planet (a precise
  term which can be properly computed from the more exact
  Equation~\ref{massfunction}).  For scale, we can express the
  quantities in more familiar units:

\begin{eqnarray}
K  = &\phantom{0.}28\phantom{.} {\rm m/s} &  (1-e^2)^{-\frac{1}{2}} (P/{\rm yr\phantom{a}})^\frac{1}{3}
        (M_*/M_\odot)^{-\frac{2}{3}} (M_{\rm 
        planet}/M_{\rm Jupiter}) \sin{i} \\
\phantom{K} = &\phantom{.}200\phantom{.} {\rm m/s} &  (1-e^2)^{-\frac{1}{2}} (P/{\rm day})^\frac{1}{3}
        (M_*/M_\odot)^{-\frac{2}{3}} (M_{\rm 
        planet}/M_{\rm Jupiter}) \sin{i} \\
\phantom{K} = &0.09\phantom{.} {\rm m/s} &  (1-e^2)^{-\frac{1}{2}} (P/{\rm yr\phantom{a}})^\frac{1}{3}
        (M_*/M_\odot)^{-\frac{2}{3}} (M_{\rm 
        planet}/M_{\rm Earth}) \sin{i} \\
\phantom{K} = &0.64 \phantom{.}{\rm m/s} &  (1-e^2)^{-\frac{1}{2}} (P/{\rm day})^\frac{1}{3}
        (M_*/M_\odot)^{-\frac{2}{3}} (M_{\rm 
        planet}/M_{\rm Earth}) \sin{i} 
\end{eqnarray}

We see here why the first strong exoplanet detections
\citep{Latham89,Mayor_queloz} were of Jovian planets in short
period orbits: 51 Peg {\it b}, for instance, has minimum mass of $0.4
M_{\rm Jupiter}$ and a 4.2 day orbit, so its RV amplitude is a
relatively large 60 m/s.  Jupiter induces a 12 m/s amplitude motion on
the Sun; the Earth's motion is a (currently) undetectable 9 cm/s.

Finally, the dependence on stellar mass means that exoplanet detection
is in principle most sensitive around the lowest mass stars, a point we shall
revisit later.

\subsection{Redshift Measurements}

Starlight is imprinted with many \hbindex{absorption lines} by ions,
atoms, and molecules in stellar atmospheres, and atomic physics allows
us to calculate or measure the rest wavelengths $\lambda_{\rm rest}$ of these lines.  A
star's radial motion causes these lines to be Doppler shifted to their
observed wavelengths $\lambda_{\rm obs}$.  The Doppler formula then
allows astronomers to calculate the relative radial speed between the star and
the observatory that measured the light $v_r$ via the \hbindex{redshift} $z$ :
\begin{equation}
z \equiv  \frac{\lambda_{\rm obs} - \lambda_{\rm rest}}{\lambda_{\rm
    rest}} = \frac{1}{\gamma(1+v_r/c)}-1
\label{Doppler}
\end{equation}
\noindent where here (unlike above) $\gamma$ is the relativistic factor $1/(1-(v/c)^2)$,
$c$ is the speed of light, and $v$ is the scalar relative speed
between the frame of the star and the observatory (which is not necessarily in
the radial direction).

Radial velocity measurements made with respect to the ``laboratory''
in this manner are called \hbindex{absolute radial velocities}, and
form the basis of our understanding of the dynamics of the Galaxy and
the expansion of the Universe. Such measurements have {\it accuracy} limited by the
wavelength calibration of the spectrograph and understanding of
complicating factors such as the internal motions of the emitting
material and redshifts from General Relativity. Typical accuracies of
absolute radial velocities of stars are of order 100 m/s \citep{Chubak12}.

More {\it precise} measurements can be made by measuring
\hbindex{differential radial velocities}, that is, the change in the
redshift between two epochs.  Differential measurements have the
advantage that some uncertainties (such as those from systematic
effects, imperfectly known rest wavelengths, or the model of the
emitting material) will effect all measurements made with a certain
instrument or technique equally, and so {\it differences} between
measurements do not suffer from them.  Thus, measurements of the {\it
  change} in the redshift of a star's spectral features can be made to
more than two orders of magnitude better precision than the accuracy
of the absolute
redshift.

\subsection{Barycentric Motion}

The observatory measuring a radial velocity is on a moving platform:
the Earth. The Earth rotates at $\sim 300$ m/s and orbits the Sun at
$\sim 30$ km/s.  As a result, the measured radial velocity of a
perfectly stable star will appear to vary on diurnal and annual
timescales by some fraction of these amounts. Correcting measurements
for this motion is known as a \hbindex{barycentric correction}, where
the term {\it \hbindex{barycenter}} refers to the center of mass of
the Solar System.  The idea is that one must correct measurements made
on the Earth to measurements that {\it would} have been made in the
(inertial) frame comoving with the Solar System's center of mass.

These motions are often much larger than the orbital motions of the
stars, and in the case of stars moving under the influence of {\it
  planets} the barycentric correction can be four or five orders of
magnitude larger. Thus, practically speaking, the problem of radial velocity precision is not
one of measuring very small redshifts, but one of measuring modest
redshifts very precisely, often to one part in $10^4$ or better.

Fortunately, the motion of the Earth is well studied and
well measured for purposes that require much more precision than
exoplanet detection.  \hbindex{Ephemerides} (tables describing the
location of celestial bodies as a function of time, pronounced {\it
  eff-em-AIR-i-deez}, singular {\it ephemeris}, pronounced {\it eh-FEM-er-iss}) for the Earth in
the barycentric frame are maintained by the Jet Propulsion Laboratory
and others, and the orientation of the Earth can be predicted into the
future with high accuracy (and measurements of that orientation are
available through, for instance, the International Earth Rotation
Service). Finally, all modern observatories provide precise
timekeeping so observations can be tagged with an appropriate
time stamp for barycentric correction algorithms later.  A detailed
description of the barycentric correction process in the context of
exoplanet detection can be found in \citet{bary}.

\section{Measuring Precise Radial Velocities}

Two primary methods of precise \hbindex{radial velocimetry} have been
used to discover and characterize exoplanets via the reflex
velocities of their host stars: absorption cell spectroscopy and
spectrograph stabilization.  They differ primarily in the method used
to calibrate the spectrograph.

The chief limit to the precision of differential redshift measurements
is the wavelength calibration of the spectrograph. Typical
RV spectrographs resolve starlight with a power of $R = \lambda / \Delta \lambda
\sim$ 50,000 --
100,000, meaning that a shift of a single pixel corresponds to a
change in radial velocity of 1 km/s. Since giant planets change their
host stars's velocities by of order tens of m/s, one must measure
shifts to a precision of $10^{-2}$ pixel for giant planets and two or
three orders of magnitude better than that to detect Earth-mass
planets. Since typical pixels on astronomical detectors are of
order 15 $\mu$m across, one is measuring the ``motions'' of stellar
lines to a fraction of a nanometer.  

Astronomical spectrographs are not this stable.  Most are general-purpose instruments with many moving parts that are actuated so the
spectrograph can be used for many purposes.  A given wavelength of
light cannot typically be expected to land to its usual position by
much better than a pixel from night to night (or year to year).  The
solution then is to employ some combination of stabilization and
calibration and employ differential
techniques.

\citet{Wright_Gaudi} provide a discussion of the historical development and first
applications of both techniques, and \citet{EPRVII} provides an
overview of the state of  the field.

\subsection{Stable spectrographs}

The first strong exoplanet detection (recognized as such only after
the fact) was that of \citet{Latham89}, who stabilized a spectrograph
in two ways: by removing it from the telescope (to prevent its
orientation with respect to the local gravity vector from changing,
minimizing flexure) and coupling it to the starlight via an optical
fiber (which served to ``scramble'' the starlight, presenting the
spectrograph with a uniform image of the star despite variations in
guiding and seeing).  Remaining variations in the wavelength solution
of the spectrograph (from
thermal changes to the spectrograph or small variations in the fiber
illumination) were tracked with a thorium-argon emission lamp, which
provided a stable set of reference lines. This combination of stabilization and
calibration allowed for precise differential measurements over
the course of several nights

Since then, this technique has been taken to extreme lengths.  State-of-the-art stable radial velocimeters today control the vibration,
temperature, and pressure of spectrographs with exquisite precision
using cryostats and vacuum chambers.  The remaining, unavoidable
changes in the spectrograph (from, for instance, slow changes in the crystalline structure
of the metals involved or irregular thermal outputs from the detector
electronics) are tracked via emission sources such as laser frequency
combs, which are locked to atomic clocks and provide essentially
perfect wavelength references.

Today, the state of the art is represented by the HARPS \citep{Queloz01_HARPS} and ESPRESSO \citep{ESPRESSO} spectrographs of
ESO, which are stable below the 1 m/s level (the latter aspires to
10 cm/s precision.)

\subsection{Absorption cell spectroscopy}

A more widely applicable method, and the one responsible for most of the
first several dozen exoplanet discoveries, is that of absorption cell
calibration.  A cell of gas is placed in the path of the starlight,
imprinting it with a set of spectral absorption features.  These
features follow the starlight through the spectrograph, and so suffer
all of the same instrumental shifts as the starlight itself.  They
thus provide an opportunity to track and calibrate all changes in a
spectrograph, no matter how unstable.

Early work by \citet{Campbell79} used HF gas and achieved precision
near 10 m/s.  Later, \citet{Marcy92} found success with molecular
iodine (I$_2$), and \citet{Butler96b} demonstrated precision near 3 m/s
through careful modeling of the iodine spectrum and the spectrograph
line spread function.  Many high-resolution spectrographs have since
been retrofitted with absorption cells, turning them into 
Doppler velocimeters capable of 1--10 m/s precision.  

\section{Radial Velocity Jitter: Spurious Doppler Signals}

Observed stellar spectroscopic absorption features are the product of
absorption, emission, scattering, and motions of gasses throughout
stellar atmospheres, across the differentially rotating stellar disk.
The lines will change their shape and centroid positions due to many
effects besides a true center-of-mass movement of the star
itself. This effect is called \hbindex{jitter} and it operates on a variety
of timescales, amplitudes, and with a range of noise distributions.

The problem is most severe for spotty, rapidly rotating, and
low-gravity stars.  This not only makes detection sensitivity
progressively worse for younger and more evolved stars, but also introduces
the more insidious problem of 
{\it false} detections around these stars, a problem that has plagued
the field since its inception \citep[e.g.][]{Queloz01}.  

The solution to the problem of detecting
planets with Doppler amplitudes near or below the level of the jitter
requires a variety of avoidance and mitigation techniques.

\subsection{Stellar Magnetic Activity}

Cool stars---those with convective envelopes and enough spectral
features that precise Doppler work is possible---have dynamos that
generate surface magnetic fields.  These fields interact strongly with
the stellar atmosphere and wind, which are (at least partly) ionized,
and affect their motions. 

The wind in particular is heavily influenced
by the field, which is anchored to the rotating star, and this
coupling causes the wind to carry angular momentum away from the star,
which thus spins down.  This spin-down weakens the dynamo, lessening
the effects of activity-based jitter.  Young stars thus spin quickly
and have a lot of activity-based jitter, and the problem is not as
severe for older stars.

On the surface of stars, magnetic fields cause surface brightness
inhomogeneities, including bright plage and faculae, and dark spots.
This breaks the symmetry of the rotational broadening kernel of the
emergent intensity from the stellar atmosphere, producing asymmetric
line profiles. That is, a bright spot on the approaching limb of the
star will produce a line with a slightly blueshifted centroid with
respect to a homogeneous stellar surface, and a spot there will produce a
redshifted line.  This effect is proportional to the spot contrast and
surface coverage, and also the projected rotational velocity of the
star, both of which are worse for younger stars.

\subsection{Photospheric Motions and Global Oscillations}

Cool stars have surfaces characterized by \hbindex{granulation}---a network of
cells with centers of hot, rising material and edges of cooler,
sinking material.  The rising material (moving towards the observer) is hotter, and so
contributes more light to the spectrum, resulting in a
\hbindex{convective blueshift}.  The velocities at different heights
in the atmosphere, the different angles taken by our line of sight at
different parts of the stellar disk, and the turbulent nature of the
motions combine produce an asymmetric line profile, overall. Finally,
the stochastic creation, destruction, and reorganization of these 
granules make this profile time-variable.  

The amplitudes of these
variations are of order meters per second, and decrease with
increasing surface
gravity \citet{Bastien13,Bastien14}.  These motions are also altered
by the strength of the global surface field strength, resulting in RV
variations on the timescales of stellar activity cycles \citep{Lovis11}.

Stars also undergo global oscillations excited by the surface
granulation and magnetic events.  These are probed by the Sun using
helioseismology (and on stars via \hbindex{asteroseismology}), and the dominant mode
is the 5 minute p-mode oscillation.  The up and down
motions visibile on different parts of the stellar disk due to these oscillations do not
precisely cancel, and the result is precise radial velocity variations
on minutes-to-hours timescales with amplitudes of order 1 m/s.
\citep{Kjeldsen05} 

\subsection{Identifying Jitter}

There are two broad strategies for dealing with jitter: avoidance and mitigation.

Avoidance is a viable strategy for surveys that can seek the
``quietest'' stars, especially those that require no more than 1 m/s
precision.  In this case, bright, unevolved, old, late-G and K dwarfs
offer the best opportunity to find low-amplitude planets with minimal
jitter mitigation. \citep[e.g.][]{Wright05,Howard09}.

Below 1 m/s, or for other kinds of stars, one must deal with jitter.
The different kinds of jitter require different mitigation
strategies.  The two most common are to correlate radial velocities with activity
indicators, such as emission in the cores of deep lines and
photometry, and to carefully examine the shapes of lines for evidence
of non-center-of-mass line shifts.

\subsubsection{Magnetic Activity Indicators}

Stellar magnetic activity results in strong, tangled field lines in
stellar atmospheres, which heat the gas via
magnetic reconnection and Alfv\'en waves.  This creates the
temperature inversion responsible for the existence of stellar
chromospheres, which, being optically thin in the continuum, cool
predominantly via emission in resonance lines such as Ca{\sc ii} H \&
K, Na{\sc i} D, and H-$\alpha$.  This emission appears in stellar
spectra as a filling-in or inversion of the cores of these absorption
features in cool stars. 

The amount of emission in the cores of these lines gives an indication
of the amount of cooling (and, therefore, heating) in the
chromosphere, and so serves as a good proxy for the overall level of
magnetic activity on the star (at least, on the hemisphere facing the
Earth).  

Perhaps surprisingly, given the somewhat tenuous connection between
the overall strength of the global field and the mechanisms by which
that field creates spurious RV signatures, the measured strength of these
emission features often has a simple relationship to the Doppler
anomaly.  

In the case of rotationally modulated spots, the Doppler
signature is out of phase with the activity measurements by $\pi$/2, because the
spots are most prominent at the center of the disk, where they have
zero rotational motion in the radial direction. In this case,
photometry may also show a signal in-phase with the activity
measurements \citep{Queloz01}.

In the case of magnetic activity cycles and other variations in the
global field strength, the effect is a simple correlation, with
stronger fields yielding more redshifted lines.  Neglecting to check
for such a correlation can lead to spurious planet detections
\citep{Wright08,Robertson14}.  In some cases, this simple correlation is probably due
to the suppression of convective blueshift as the field lines restrict the motion of
the (slightly) ionized atmosphere \citep[e.g.\ as in the case of
$\alpha$ Cen B][]{AlphaCenBb}.

Even in the case where a direct correlation between activity
indicators and RVs is not clear, the effects of stellar magnetic
activity can be diagnosed via their rotational modulation.  For
sufficiently densely sampled activity time series, a power spectrum will often
reveal the rotation period (and harmonics thereof).  Any RV variation
at these periods is suspect.  Of course, real planets may have orbital
periods that match these periods \citep[as in the case of
$\tau$ Boo {\it b}][]{Butler97}, whether simply by coincidence or due to
tidal locking, but in general the burden of proof for a claimed
exoplanet discovery grows higher in the
presence of such coincidences, especially for low-amplitude planets \citep{Robertson14b}.

\subsubsection{Line shape analysis}

Sources of spurious RV changes will, in general, not have
exactly the same spectral signature as a Doppler shift due to true
center-of-mass motion.  While true Doppler shifts will leave line
shapes unchanged, rotationally modulated spots or changes to the
convective blueshift pattern will alter those shapes, if only
slightly.  

The most commonly seen measures of line shape are the line
bisector---which traces the center a line as a function of depth below
the continuum---and the width of the lines.  True Doppler shifts will
preserve these shapes, while 
changes in a line's profile should alter them \citep{Hatzes96}.  The ``span''
of the bisector is the difference in the bisector position near the top and
bottom of the line, measured in units of velocity.  This is equivalent
to the inverse of the mean slope of the bisector (for continuum
normalized spectra), and is usually designated BIS \citep{Queloz01,Boisse11}.

Correlations between measured radial velocities and BIS or line widths are thus red
flags that the signal is not due to the reflex motion of an orbiting
companion.  It may indicate a spectrum blended with light from other
stars, stellar pulsations, or other sources of jitter \citep[see][for
several examples.]{Wright13}  

Bisector and other line shape variations are measured most easily in
stable spectrographs, where the shape of the cross-correlation function (CCF) can
serve as a proxy for the mean shape of all lines used to derive
velocities.  They are not routinely measured in absorption cell spectroscopy.

\subsubsection{Wavelength Dependence}

True center-of-mass motion shifts all line
wavelengths by the same fraction, while spurious effects will
generally have different effects on lines of different wavelengths,
depths, formation heights, and ionization states.  For instance,
rotationally modulated spot-induced RV anomalies should be less
pronounced in the infrared, where brightness temperature contrasts are
lower.

In principle, RVs measured from various lines should be checked for
consistency, but in practice this is difficult.  In absorption cell
velocimetry, the spectrum is not decomposed into individual lines, and
only a relatively narrow range of wavelengths is examined, so measured
RVs cannot be examined as a function of wavelength or line
properties.  The problem is more tractable in stable spectrograph
velocimetry, where individual lines are typically chosen for the
analysis and can, in principle, be compared with each other for consistency.

\subsection{Target Selection}

The first targets of precise Doppler surveys were generally very
bright (naked-eye) stars because early practitioners were generally
using small telescopes, and because of the large number of photons
needed to make a precise Doppler measurement. The stars must also have
a rich set of absorption features to measure, making hot stars
unsuitable targets, and must have narrow features, favoring stars with
low projected rotational velocities ($v \sin{i} <10$ km/s).  Young stars have many
surface features and flares that make them unsuitable for the highest
precision work, in addition to their high rotational velocities. 

Giant stars also turn out to be poorer precise Doppler targets,
because their atmospheres exhibit large variability in their
Doppler motions \citep[e.g.][]{Hekker06}.  The highest precision is
thus achieved on bright, old, dwarf, cool stars, which typically show
intrinsic variations only at the 1 m/s level \citep{Wright05}.

There are many reasons to push the technique outside of this optimal
region.  One, is exploration of the dependence of planet occurrence on stellar
mass \citep[e.g.][]{Johnson10b}.  Another is to measure the masses of
planets known to transit, such as the faint, sometimes young or
evolved targets of the {\it Kepler} and {\it K2} missions
\citep{Kepler}.  

\subsection{Optical vs. Infrared}
\label{infrared}

In general, precise Doppler work is most easily performed in the
optical, where absorption cells are well calibrated, detectors are
better behaved and more easily calibrated, and telluric absorption and
emission is less difficult to work with. There are compelling reasons
to pursue infrared precise Doppler work, however.  

One is if one wishes to optimize not precision, but sensitivity
to planets in the Habitable Zone \citep{Kasting93}, where surface
liquid water is most likely to be found.  Very cool late-M dwarfs have
close-in Habitable Zones, and so these planets have shorter
periods than around G and K dwarfs.  This, combined with the more favorable
planet-to-star mass ratios for a given planet mass, makes the necessary
precision for discovering their Habitable-Zone planets an order of
magnitude less stringent.  Since these very cool stars have very
little optical luminosity, the only practical solution is to measure
their spectra in the near infrared where they have most of their
energy output.

Recently, these and other concerns have created a strong push to
extend precise Doppler velocimetry to the near infrared, with stable
instruments such as the Habitable Zone Planet Finder and CARMENES
\citep{HPF,CARMENES}; and gas cell work as well \citep{Bean10,Gao16}.

\section{Future Challenges and Opportunities}

Precise radial velocimetry will continue to be a cornerstone of
exoplanetary research into the foreseeable future.  Space-based
transit surveys are often limited by the availability of radial
velocities to rule out many common sources of false positives, and to
measure the masses of the planets discovered.  Discovery of Jupiter
analogs and very-long-period companions require decades of
observation, making archival radial velocities relevant for years to
come.   

Advances in both instrumental precision and the understanding of the
sources of and methods for mitigating stellar jitter will push radial
velocity sensitivity down to lower and lower mass planets, with a
common goal being the detections of an Earth analog (requiring
detection of a 9 cm/s signal).  The most ambitious plans call for
stability of 1 cm/s with spectrographs that can access enough
collecting area to ensure the measurements are not photon limited
\citep[e.g. CODEX][]{Pepe08}.


\bibliographystyle{spbasicHBexo}  

\end{document}